\documentclass[sigconf, nonacm]{acmart}

\usepackage{multirow}
\usepackage{color}
\usepackage{listings}
\usepackage{threeparttable}

\usepackage{enumitem}
\usepackage{xcolor}
\usepackage{balance}
\lstdefinestyle{sqlred}{
    language=SQL,
    basicstyle=\ttfamily\small,
    keywordstyle=\color{black},
    identifierstyle=\color{black},
    commentstyle=\color{gray},
    moredelim=[is][\color{red}]{<r>}{</r>},
    breaklines=true,
    columns=fullflexible
}

\usepackage[most]{tcolorbox}
\usepackage{xcolor}
\usepackage[normalem]{ulem} 

\definecolor{highlightred}{HTML}{D62728}     
\definecolor{boxbg}{rgb}{0.98,0.98,0.98}     
\definecolor{boxframe}{rgb}{0.80,0.80,0.80}  
\definecolor{headerbg}{HTML}{D2D9F2}  
\definecolor{headergreen}{HTML}{B7D7B0} 
\definecolor{symbolcolor}{HTML}{0050B0} 
\definecolor{exampleheader}{HTML}{D2D9F2} 
\definecolor{failureheader}{HTML}{F2C6C2} 
\definecolor{failureheader2}{HTML}{D88C8C} 

\tcbset{
  sqlbox/.style={
    colback=gray!10,
    coltitle=black,
    colbacktitle=exampleheader,
    sharp corners,
    fonttitle=\bfseries,
    boxrule=0.3pt,
    top=3pt, bottom=2pt,
    left=4pt, right=4pt,
    fontupper=\small,
    fontlower=\small,
    enhanced,
    boxsep=3pt,
    title filled
  }
}

\tcbset{
  examplebox/.style={
    colback=gray!10,
    coltitle=black,
    colbacktitle=exampleheader,
    boxrule=0.4pt,
    left=6pt, right=6pt, top=4pt, bottom=4pt,
    fontupper=\small,
    fontlower=\small,
    enhanced,
    breakable
  }
}

\newcommand{\symb}[1]{\textcolor{blue}{\boldmath{$#1$}}}
\newcommand{\sqlkw}[1]{\textcolor{blue}{#1}}

\tcbset{
  promptbox/.style={
    colback=gray!10,
    coltitle=black,
    colbacktitle=headergreen,
    fonttitle=\bfseries,
    sharp corners,
    boxrule=0.4pt,
    top=4pt, bottom=3pt,
    left=6pt, right=6pt,
    fontupper=\small,
    fontlower=\small,
    enhanced,
    breakable,
    title filled
  }
}

\newcommand\vldbdoi{10.14778/3797919.3797935}
\newcommand\vldbpages{1291 - 1304}
\newcommand\vldbvolume{19}
\newcommand\vldbissue{6}
\newcommand\vldbyear{2026}
\newcommand\vldbauthors{\authors}
\newcommand\vldbtitle{\shorttitle} 
\newcommand\vldbavailabilityurl{https://github.com/mmdrez4/SQL-Exchange}
\newcommand\vldbpagestyle{empty} 

\begin{document}
\title{SQL-Exchange: Transforming SQL Queries Across Domains}

\author{Mohammadreza Daviran}
\affiliation{%
  \institution{University of Alberta}
  \city{Edmonton}
  \state{AB}
  \country{Canada}
}
\email{daviran@ualberta.ca}

\author{Brian Lin}
\affiliation{%
  \institution{University of Alberta}
  \city{Edmonton}
  \state{AB}
  \country{Canada}
}
\email{shihhsu1@ualberta.ca}

\author{Davood Rafiei}
\affiliation{%
  \institution{University of Alberta}
  \city{Edmonton}
  \state{AB}
  \country{Canada}
}
\email{drafiei@ualberta.ca}

\begin{abstract}
We introduce SQL-Exchange, a framework for mapping SQL queries across different database schemas by preserving the source query structure while adapting domain-specific elements to align with the target schema.
We investigate the conditions under which such mappings are feasible and beneficial, and examine their impact on enhancing the in-context learning performance of text-to-SQL systems as a downstream task.
Our comprehensive evaluation across multiple model families and benchmark datasets---assessing structural alignment with source queries, execution validity on target databases, and semantic correctness---demonstrates that SQL-Exchange is effective across a wide range of schemas and query types. Our results further show that both in-context prompting with mapped queries and fine-tuning on mapped data consistently yield higher text-to-SQL performance than using examples drawn directly from the source schema. 

\end{abstract}

\maketitle

\pagestyle{\vldbpagestyle}
\begingroup\small\noindent\raggedright\textbf{PVLDB Reference Format:}\\
\vldbauthors. \vldbtitle. PVLDB, \vldbvolume(\vldbissue): \vldbpages, \vldbyear.\\
\href{https://doi.org/\vldbdoi}{doi:\vldbdoi}
\endgroup
\begingroup
\renewcommand\thefootnote{}\footnote{\noindent
This work is licensed under the Creative Commons BY-NC-ND 4.0 International License. Visit \url{https://creativecommons.org/licenses/by-nc-nd/4.0/} to view a copy of this license. For any use beyond those covered by this license, obtain permission by emailing \href{mailto:info@vldb.org}{info@vldb.org}. Copyright is held by the owner/author(s). Publication rights licensed to the VLDB Endowment. \\
\raggedright Proceedings of the VLDB Endowment, Vol. \vldbvolume, No. \vldbissue\ %
ISSN 2150-8097. \\
\href{https://doi.org/\vldbdoi}{doi:\vldbdoi} \\
}\addtocounter{footnote}{-1}\endgroup

\ifdefempty{\vldbavailabilityurl}{}{
\begingroup\small\noindent\raggedright\textbf{PVLDB Artifact Availability:}\\
The source code, data, and/or other artifacts have been made available at \url{\vldbavailabilityurl}.
\endgroup
}

\section{Introduction}
What do \textit{toxicology} and \textit{Formula 1} have in common, and where do \textit{superheroes} intersect with \textit{student clubs}? These are all databases in the BIRD benchmark~\cite{li2023ca}, and despite the differences in the semantics of their tables and columns, the structure of their queries is similar, and in many cases, identical. Consider the following example queries across three different databases:
\begin{enumerate}
    \item \textsf{What is the total number of clients in the Vsetin district?}
    \item \textsf{What is the total number of atoms in molecules with a label of '+'?}
    \item \textsf{What is the total number of races held at the Canadian Grand Prix circuit?}
\end{enumerate}

\begin{figure*}[t]
  \centering
    \includegraphics[width=\textwidth]{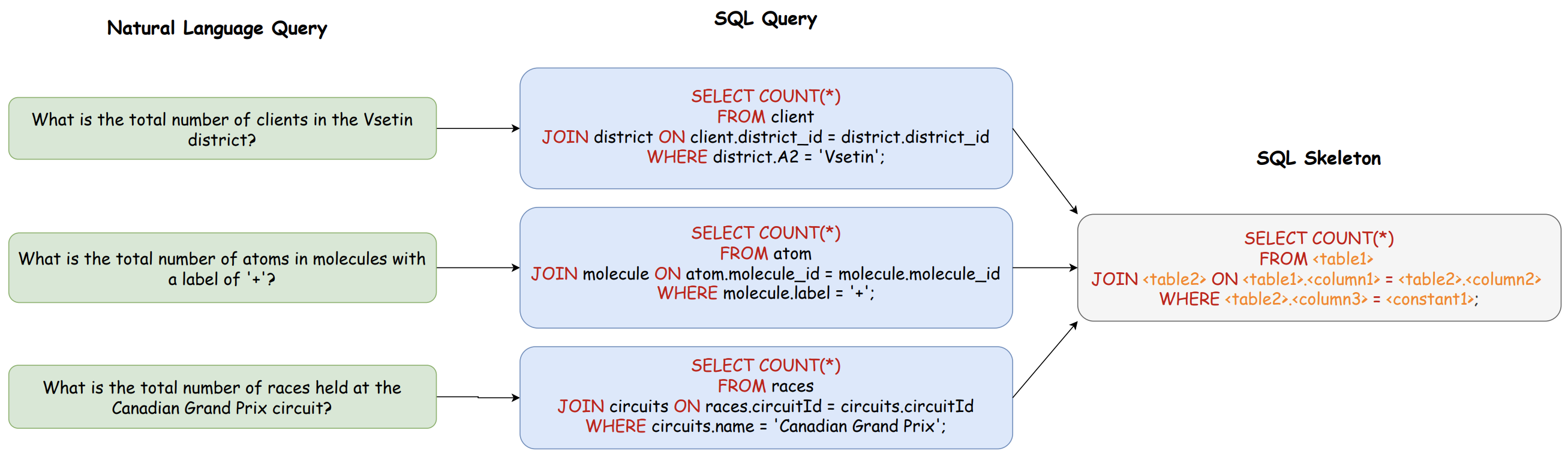}
    \Description{Three natural language questions are shown alongside their corresponding SQL queries. Each query uses a SELECT COUNT and a JOIN with a WHERE condition. All three share the same SQL skeleton: SELECT COUNT(*) FROM <table1> JOIN <table2> ON <table1.column1> = <table2.column2> WHERE <table2.column3> = <constant1>.}
  \caption {Illustration of three natural language queries, their corresponding SQL translations, and their shared SQL skeleton, demonstrating structural similarity across different database schemas.}
    \label{fig:query_structure}
\end{figure*}

As shown in Figure~\ref{fig:query_structure}, all three queries share the same SQL skeleton despite the differences in their query semantics and schema links. The question studied in this paper is if queries expressed on a source database can be mapped to equivalent queries on a target database. Here, we use the term `equivalence' loosely to refer to queries that are structurally identical, independent of schema links and constants. 



Although the source and target domains may differ semantically, structurally aligned queries are valuable in several practical scenarios. Many real databases, especially newly created or under-documented ones, lack sufficient NL–SQL pairs for in-context learning or fine-tuning, yet structurally relevant examples are known to significantly improve performance; for example, \citet{pourreza2024sql} reports that using in-context examples tailored for the target domain improves SQL generation performance by 12\% compared to using cross-domain examples. SQL-Exchange enables the automatic creation of schema-aligned SQL–NL pairs directly for the target schema, supporting prompting, data augmentation, and model adaptation in low-resource settings. Structural mappings are also useful for bootstrapping query repositories: in the SDSS project~\citep{sdss}, a repository of queries was constructed for a new database to support the search and reuse of similar queries~\citep{howe2011automatic,khoussainova2011session}. Additionally, educational and assessment settings~\citep{meissner2024evalquiz,jury2024evaluating} benefit from structurally parallel queries, where varying the domain while preserving logical structure reduces plagiarism and helps maintain consistent difficulty across assignments.





Existing work on data migration---an industry valued at USD 8.2 billion in 2021 and projected to reach USD 33.58 billion by 2030~\citep{dataMigMarket}---has traditionally focused on transforming data and queries across different database engines and architectures (e.g., SQL-to-NoSQL, SQL Server to SQLite, and SQL to Hive~\cite{wang2015qmapper}).
In contrast, text-to-SQL research predominantly focuses on translating natural language into SQL under a single, fixed schema~\citep{li2023resdsql,pourreza2024din,gao2024text}.
Parallel efforts explore other query translation tasks, such as SQL-to-English~\citep{luk1986elfs}, and SQL-to-NoSQL~\citep{chung2014jackhare}, aimed at improving accessibility or enabling cross-platform  execution. However, most prior work focuses on syntactic translation or query generation rather than preserving the structural logic of SQL queries while adapting them to new database schemas. To the best of our knowledge, no existing work systematically addresses schema-level SQL query translation---that is, mapping queries to databases with different schemas while preserving their structural integrity. We address this gap with a novel LLM-based framework that maps SQL queries across schemas, maintaining the original query structure and enabling seamless transfer of logic between databases.

There are intriguing questions and challenges in mapping queries across domains. 
First, queries that are relevant in one domain may not be meaningful in another due to differences in the number and semantics of tables and columns, as well as the presence or absence of foreign key relationships and constraints.
Second, as query complexity increases, they may involve multiple joins, nested subqueries, aggregations, and filtering conditions, all of which introduce dependencies that must be correctly adapted to the target schema, making it more difficult to transfer the query logic to a different domain. Furthermore, the relationships between tables and columns in different schemas may be structurally different, requiring transformations that go beyond simple column substitutions. The impact of these factors on query portability is neither well-studied nor well-understood.





In this paper, we (1) introduce a novel approach for mapping SQL queries across different domains, while preserving their logical structure, and (2) conduct a comprehensive empirical investigation into its behavior and limitations.
This work aims to address three key questions: \textbf{(1)} Can SQL queries be accurately mapped across different database schemas while preserving their structural form and logic?
\textbf{(2)} Can LLMs serve as reliable tools for performing such mappings?  
 and \textbf{(3)} Do mapped queries effectively improve the performance in downstream tasks such as text-to-SQL generation?  

To answer these questions, we undertake an extensive, multi-dimensional evaluation that characterizes the structural fidelity, execution validity, and natural-language meaningfulness of the mapped queries across BIRD and SPIDER. Our analysis reveals that SQL-NL pairs generated through SQL-Exchange exhibit strong structural alignment (over 82\%) and high NL quality (over 95\%) using Gemini-1.5-flash, with GPT-4o-mini showing similarly strong trends.
These findings are supported by small-scale manual evaluations and larger-scale assessments using LLM-as-a-judge. 





Our contributions can be summarized as follows:  
\textbf{(1)} We introduce and analyze a framework for cross-domain SQL query mapping that preserves the structural logic of a source query while adapting its schema-dependent components to a target database.  
\textbf{(2)} Through extensive evaluation on BIRD and SPIDER using multiple LLM families, we show that SQL-Exchange produces queries that remain structurally faithful, executable, and semantically meaningful. We also identify the main patterns of structural adaptation and the key factors that affect mapping quality.    
\textbf{(3)} We construct a corpus of over 100{,}000 schema-aligned SQL–NL pairs and demonstrate their downstream utility for \textit{text-to-SQL}: SQL-Exchange examples consistently outperform unmapped and zero-shot baselines across retrieval methods, and fine-tuning on mapped examples yields substantially larger gains than fine-tuning on source-only data, underscoring the method’s value in low-supervision settings.

\section{Related Work}

\paragraph{\bf Synthetic Data Generation with LLMs}

LLMs have been extensively used for synthetic data generation across various natural language processing tasks, including classification, data augmentation, and code generation~\citep{yoo-etal-2021-gpt3mix-leveraging, li-etal-2023-synthetic, dai2025auggpt, nadas2025synthetic}. 
These efforts typically aim to expand the volume and diversity of training data by generating paraphrases or novel examples under controlled conditions.
A particular relevant area of application is question generation over structured knowledge bases and narrative texts~\citep{yoon-bak-2023-diversity, liang-etal-2023-prompting, li-zhang-2024-planning}. For instance, \citet{li-zhang-2024-planning} introduce a planning-first strategy that guides LLMs in generating questions with greater control over both structure and content.
Most of these methods focus on generating isolated, free-form questions with minimal structural constraints---significantly different from the more complex task of generating SQL–NL pairs, which requires maintaining consistency between natural language and formal query logic.

\paragraph{\bf Synthetic SQL Generation and Cross-Dialect Translation}

Early work on synthetic data generation for text-to-SQL employed template-based SQL synthesis and copy-based Seq2Seq models to translate SQL into natural language~\citep{guo-etal-2018-question, wu-etal-2021-data, wang-etal-2021-learning-synthesize, hu-etal-2023-importance}. More recent approaches leverage LLMs to generate SQL–NL pairs grounded in real database schemas~\citep{lupidi2024source2synth,pourreza2024sqlgen}. Large-scale pipelines such as OmniSQL~\citep{li2025omnisql}, SQL-Factory~\citep{li2025sql}, and SQLForge~\citep{guo-etal-2025-sqlforge} further expand training corpora through multi-agent generation, AST-level composition, and broad schema exploration.

These systems, however, are designed for data expansion: they generate new SQL queries—often across thousands of synthetic or isolated schemas—to support large-scale pretraining. In contrast, SQL-Exchange does not aim to enlarge training sets or synthesize new schemas. Instead, it transforms an existing SQL query into another schema while preserving its structural scaffold. This enables domain-adapted one-shot or few-shot in-context examples, a capability not supported by synthetic generation pipelines whose outputs are not tied to the logical structure of a specific source query. Whereas synthetic generators optimize for coverage and diversity, SQL-Exchange preserves cross-domain structural alignment.




\newpage

A parallel line of work adresses SQL dialect or engine translation~\citep{ngom2024mallet, zmigrod2024translating, lin2024momq}, using rule-based methods, traditional models, or LLM-assisted pipeline to improve query portability. 
These approaches typically assume identical schemas across dialects or target structurally simple SQL queries, limiting their applicability to schema-adapted transformations.

\paragraph{\bf Schema Mapping and Data Migration}
Data migration---the process of moving data between formats, systems, or platforms---has a long history in the database and systems communities~\cite{hull1984relative,miller1993use,khuller2003algorithms,lu2002aqueduct}. When data spans multiple databases with differing schema representations, a central challenge is managing semantic heterogeneity across database schemas~\cite{hull1984relative,hull1997managing}.

Schema conversion is often employed in SQL-to-NoSQL migrations, particularly to improve join performance by reorganizing data so that join-relevant attributes reside in the same document or collection~\cite{zhao2014schema}. In NoSQL systems, schema evolution may occur lazily, driven by application interactions and without requiring downtime or disrupting active workloads~\cite{saur2016evolving}. Query transformations are also frequently necessary; for example, translating SQL to Hive often requires modifying queries when certain SQL constructs are not supported by the target system~\cite{wang2015qmapper}.

Unlike traditional data migration techniques, which typically preserve the original data and query semantics despite changes in representation or syntax, cross-domain query transformation---as considered in this work---allows both data values and query intent to shift. In our setting, the source query serves as a structural guide rather than a strict equivalence, enabling adaptation to target schemas with differing semantics or domain-specific conventions.

\section{Methodology}
We aim to translate a natural language query and its corresponding SQL from a source schema to an equivalent query pair on a target schema while preserving the original logical structure.

\subsection{Problem Formulation}

Let $q_s$ be a natural language question and $sq_s$ its corresponding SQL query over a source schema $s$. Given a target schema $t$, our goal is to generate a natural language question $q_t$ and an SQL query $sq_t$ over $t$ such that:
\begin{itemize}
    \item $q_t$ is a meaningful and well-formed question over $t$;
    \item $sq_t$ correctly expresses the semantics of $q_t$ within $t$;
    \item $sq_s$ and $sq_t$ share the same logical structure, abstracting away from table names, column names, and constants.
\end{itemize}


\subsection{Limitations of Zero-Shot Translation}

Large Language Models (LLMs) have demonstrated impressive capabilities in translation tasks ranging from natural language to formal representations (e.g., text-to-SQL~\citep{pourreza2024din,gao2024text}) to cross-format conversions (e.g., SQL to NoSQL~\citep{chung2014jackhare} and tabular data representations~\citep{nobari2024tabulax,huang2024transform}). To assess their suitability for schema-to-schema SQL translation, we began with a zero-shot setting in which the model received only a high-level task description and schema metadata.

Given $(q_s, sq_s, s, t)$, the LLM was asked to generate $(q_t, sq_t)$ directly. This setup revealed two critical challenges:

\textit{Structural Drift.} The model frequently failed to preserve the structural backbone of the source SQL. Structural alignment---the degree to which the high-level logic (e.g., joins, aggregations, filters) is retained---was often below 50\%, and in some benchmarks, below 15\%. A common issue was the omission of essential \texttt{JOIN} clauses, flattening multi-table logic into oversimplified single-table queries.

\textit{Schema Leakage.} The model also showed a tendency to copy table names, column names, and constants from the source schema into the target query, without adaptation. For instance, GPT-4o-mini reused 15\% of source column names and over 70\% of literal values in one dataset, often leading to syntactically incorrect or semantically irrelevant queries.

\begin{figure}[ht]
\centering

\begin{tcolorbox}[sqlbox, title=BIRD: \texttt{appstore} $\rightarrow$ \texttt{debit\_card\_specializing}]
\textbf{Source SQL:}
\begin{quote}
SELECT AVG(Price) \\
FROM playstore \\
WHERE Genres = 'Dating'
\end{quote}
\textbf{Target SQL:}
\begin{quote}
SELECT AVG(T1.\textcolor{highlightred}{Price}) \\
FROM products AS T1 \\
WHERE T1.Description = \textcolor{highlightred}{'Dating'}
\end{quote}
\end{tcolorbox}

\begin{tcolorbox}[sqlbox, title=BIRD: \texttt{soccer\_2016} $\rightarrow$ \texttt{toxicology}]
\textbf{Source SQL:}
\begin{quote}
SELECT T2.Outcome\_Type \\
FROM Match AS T1 \\
\textcolor{highlightred}{INNER JOIN Outcome AS T2} \\
\textcolor{highlightred}{ \quad ON T1.Outcome\_type = T2.Outcome\_Id} \\
WHERE T1.Match\_Id = '392195'
\end{quote}
\textbf{Target SQL:}
\begin{quote}
SELECT T1.label \\
FROM molecule AS T1 \\
WHERE T1.molecule\_id = \textcolor{highlightred}{'392195'}
\end{quote}
\end{tcolorbox}

\Description{Two examples of zero-shot SQL query mappings. In the first example, elements from the source schema, such as 'Price' and 'Dating', are incorrectly copied into the target schema where they do not exist. In the second example, the target SQL omits a required JOIN present in the source query, causing a structural mismatch.}

\caption{Examples of structural drift and schema leakage in zero-shot mapping using Gemini. In the first case, schema elements (in red)—such as column names and literals—are inappropriately copied from the source into the target query (e.g., no price column in target). In the second case, the target query omits a necessary \textsf{JOIN} clause from the source, leading to a loss of structural fidelity.}

\label{fig:zeroShot-examples}
\end{figure}

These challenges, illustrated in Figure~\ref{fig:zeroShot-examples}, motivated the development of a more structured prompting strategy, as discussed next.

\subsection{Structured Prompting with Chain-of-Thought Reasoning}


To address the weaknesses of zero-shot prompting, we adopt a structured one-shot prompting approach in which the model receives: (1) a high-level task description; (2) one illustrative example demonstrating a correct source-to-target query mapping; and (3) the source and target schema definitions together with the source query.

This format capitalizes on the benefits of in-context learning while remaining within within LLM context limits. 


Additionally, inspired by chain-of-thought (COT) prompting~\citep{wei2022chainofthought}, we incorporate an intermediate reasoning step. Rather than immediately generating $sq_t$, the model first produces a brief natural language explanation---a \emph{thought process}---that outlines how it interprets the query logic and how it plans to map it to the target schema. This step enhances interpretability and guides structurally faithful SQL generation. The complete structured prompt, including instructions and the required inputs, is shown in the box below:

\begin{tcolorbox}[promptbox, title=Structured Prompt with Reasoning]

\textbf{Input:}
A natural language question (\symb{q_s}) and SQL query (\symb{sq_s}) over a source schema (\symb{s});  
a target schema (\symb{t}); sample data from the target schema (\symb{d_t});  
and a demonstration example (\symb{e}) from another schema pair.

\textbf{Output:}
A JSON object containing a mapped natural language question (\symb{q_t}) and SQL query (\symb{sq_t}) over the target schema (\symb{t}).

\tcblower

\textbf{\#\# System prompt:} You are an expert in areas of database design and SQL queries and your job is to swap tables and columns from a query to generate a new query, and new question based on the target schema. \\

\textbf{\#\# Instructions:}

\textbf{-} The output must be in a valid JSON format.

\textbf{-} For each Source query, there must be a corresponding output in the output array.

\textbf{-} First, create \texttt{tables\_columns\_replacement} by replacing all table names, column names, and constant values in the source query with placeholders: \texttt{"table"}, \texttt{"column"}, \texttt{"constant\_value"}.

\textbf{-} Then, perform the following steps and give your thought process (in not more than 5 sentences) for each step:

  \quad \textbf{1}: Generate a new query from \texttt{tables\_columns\_replacement}.%

  \quad \quad \textbf{1.1}: Use Target schema to replace table and column names that 
  
  \quad \quad \quad \enspace \enspace make sense in terms of query.

   \quad \quad \textbf{1.2}: For constant values in \texttt{tables\_columns\_replacement}:

   \quad \quad \quad \quad  \textbf{\textbullet} Use meaningful values from the target schema sample 
   
   \quad \quad \quad \quad \enspace \space data.

   \quad \quad \quad \quad \textbf{\textbullet} Do not reuse constant values from the \texttt{"source\_query"}

   \quad \quad \quad \quad \textbf{\textbullet} Ensure that numerical constant values differ from those in 
   
   \quad \quad \quad \quad \enspace \space the \texttt{"source\_query"}.

   \quad \textbf{2}: Generate a new question based on the query that you just
   
   \quad \enspace \enspace generated. \\

  


\textbf{\#\# Example}: \{\symb{e}\} \\

\textbf{\#\# Generate the query for the following query:}\\

   \texttt{\# Source schema: \symb{s}} \\
   \texttt{\# Target schema: \symb{t}} \\
   \texttt{\# Target sample data: \symb{d_t}} \\
   \texttt{\# Source query: \{\symb{q_s}, \symb{sq_s}\}} \\

\# Output:
\end{tcolorbox}

The overall pipeline includes two key mechanisms described below: query template abstraction and constant substitution.

\subsection{Template-Guided Query Transfer}

One major failure in zero-shot settings is the model’s difficulty in disentangling schema-independent logic from schema-specific artifacts, often leading to poor structural alignment and inappropriate reuse of tables or columns from the source schema. To address this, we introduce a template-guided abstraction mechanism that helps the model disentangle logical structure from surface-level artifacts.

The process begins by converting the source SQL query $sq_s$ into a schema-agnostic \textbf{query template}, where all table names, column names, and constants are replaced with generic placeholders such as \texttt{table}, \texttt{column}, and \texttt{value}. Crucially, this abstraction preserves the structural backbone of the query—including joins, aggregations, and filters---while removing database-specific identifiers. For example, a query that joins two tables on a foreign key and applies a filtering condition becomes a high-level template with the same logical form but anonymized components. This abstraction encourages the model to focus on generalizable relational patterns rather than memorized schema tokens.

Once the template is created, the model grounds the template by instantiating it with appropriate elements from the target schema $t$. This step involves selecting valid table and column names and substituting meaningful constants, resulting in a complete SQL query $sq_t$ that aligns with both the logic of the source query and the structure of the target database. This two-phase approach promotes compositional reasoning, reduces schema leakage, and yields structurally faithful translations.

Empirical results confirm the effectiveness of this approach. On the BIRD benchmark, for example, using \textsf{Gemini-1.5-Flash}, structural alignment improved from just 13.1\% in the zero-shot setting to 66.6\% with template-guided prompting. Similar gains were observed across other models and benchmarks, demonstrating the robustness of this strategy.

To illustrate, consider this example from a schema on postal and demographic data (with structural elements highlighted in blue):
\begin{tcolorbox}[sqlbox, title=Source Schema]
\textbf{NL Question:} Provide the names of bad aliases in the city of Aguadilla. \\[0.3em]
\textbf{SQL:} \sqlkw{SELECT} T1.bad\_alias \sqlkw{FROM} avoid AS T1 \sqlkw{INNER JOIN} zip\_data AS T2 \sqlkw{ON} T1.zip\_code = T2.zip\_code \sqlkw{WHERE} T2.city = 'Aguadilla'
\end{tcolorbox}

Its abstract template becomes:
\begin{tcolorbox}[sqlbox, title=Schema-Agnostic Template]
\sqlkw{SELECT} T1.column \sqlkw{FROM} table1 AS T1 \sqlkw{INNER JOIN} table2 AS T2 \sqlkw{ON} T1.column2 = T2.column2 \sqlkw{WHERE} T2.column3 = constant\_value
\end{tcolorbox}

Within the target chemical-compounds schema, this maps to:
\begin{tcolorbox}[sqlbox, title=Target Schema]
\textbf{NL Question:} What are the bond types in molecule `TR028'?\\[0.3em]
\textbf{SQL:} \sqlkw{SELECT} T1.bond\_type \sqlkw{FROM} bond AS T1 \sqlkw{INNER JOIN} molecule AS T2 \sqlkw{ON} T1.molecule\_id = T2.molecule\_id \sqlkw{WHERE} T2.molecule\_id = 'TR028'

\end{tcolorbox}






This illustrates how structure is retained while adapting the query to an entirely different domain.




\subsection{Substituting Constant Values}
\label{sec:constant-substitution}

Substituting constant values poses a significant challenge in query mapping. Unlike table and column names, which can often be aligned through structural reasoning over the target schema, constants
---such as categorical labels (e.g., district names, molecule types) or numerical values---are highly dataset-specific and rarely transferable across domains.


To address this, we augment the prompt with a small set of sample rows from the target database. This gives the model access to valid constant values grounded in the target schema. Additionally, we explicitly instruct the LLM to replace each constant placeholder with a semantically appropriate value that exists in the target database, rather than copying constants from the source query. Without such guidance or access to target-side data, LLMs tend to default to reusing source query constants, leading to semantically invalid outputs that may be syntactically correct but inconsistent with the target schema.  This was especially evident in our zero-shot experiments, where constant reuse rates reached as high as 72\% on BIRD using \textsf{GPT-4o-mini}, indicating that most target queries failed to generate schema-appropriate values. In contrast, SQL-Exchange reduced constant reuse by more than 30 percentage points, producing values that matched the target schema distribution and yielding higher execution validity and question quality.


\section{Evaluation}
We evaluate the performance of SQL-Exchange across multiple model families, datasets, and evaluation metrics.

\subsection{Experimental Setup}




\textit{\textbf{Datasets}} We evaluate SQL-Exchange on two standard text-to-SQL benchmarks: \textit{BIRD}~\citep{li2023ca} and \textit{SPIDER}~\citep{yu-etal-2018-spider}. For each benchmark, we select one source database from the training set and one target database from the development set, applying the same instructions and prompt structure across all models and database pairs. BIRD includes \textbf{69} training and \textbf{11} development databases, while SPIDER includes \textbf{146} training and \textbf{20} development databases. We sample up to 20 queries per source database; for \textsf{GPT-4o-mini} on SPIDER, we reduce this to 10 for cost reasons. If a database has fewer than the desired number of queries, all available queries are included.

To verify that our sampling does not bias the evaluation toward simpler SQL structures, we examined the \emph{hardness} of the sampled subsets using the official Spider rule-based criteria (easy, medium, hard, extra-hard), defined by structural markers such as joins, nesting, grouping, and set operations. For BIRD, the sampled hardness distribution closely matches the full dataset, with differences below $0.6\%$ in every category; a chi-square test further confirms no statistical difference ($\chi^2 = 0.50$, $p = 0.918$). For SPIDER, the differences remain modest (below $2.8\%$ per category). Although the chi-square test detects a shift ($\chi^2 = 16.14$, $p = 0.001$), the absolute deviations are small, and the sample still contains substantial numbers of hard and extra-hard queries, preserving the overall complexity profile of the full dataset.


\textit{\textbf{LLMs for Mapping.}} For the query mapping task, we employed two LLMs from distinct model families: \textsf{GPT-4o-mini}, \textsf{Gemini-1.5-flash} \citep{gemini2024multimodal}, and the open-weight \textsf{Llama-3.3-70B}. The first two models were selected for their fast response times, cost-efficiency, and long context windows. Both offer 128k-token context windows, enabling us to batch 10 to 20 queries per prompt---alongside long schema descriptions and sample inputs---thereby reducing the number of LLM calls and minimizing token redundancy. The combined prompt length often exceeded the 16k-token context window of models such as \textsf{GPT-3.5-turbo}.  In contrast, both \textsf{GPT-4o-mini} and \textsf{Gemini-1.5-flash} handled such inputs reliably with minimal failures.
We additionally included \textsf{Llama-3.3-70B-Instruct} to evaluate SQL-Exchange under an open model.
Throughout the paper, we refer to model–dataset combinations using shorthand notations such as \texttt{BIRD–Gemini}, \texttt{BIRD–GPT}, \texttt{SPIDER–Gemini}, and \texttt{BIRD–Llama}.

\textit{\textbf{LLM for Evaluation.}} To assess semantic quality, we use \textsf{Gemini-2.0-flash} as an LLM-based evaluator due to its superior reasoning performance. This model is validated against human annotators for assessing the correctness and quality of generated queries.


\textit{\textbf{Model Access and Configuration.}} \textsf{GPT-4o-mini} was accessed through the OpenAI API,\footnote{\url{https://platform.openai.com/docs/models/gpt-4o-mini}} Gemini models via Google's developer API,\footnote{\url{https://ai.google.dev/models/gemini}} and \textsf{Llama-3.3-70B} through the Nebius API.\footnote{\url{https://tokenfactory.nebius.com/}} All models were queried with \textsf{temperature = 0.0}, \textsf{top\_p = 1.0}, and \textsf{top\_k = 0} whenever supported.


\textit{\textbf{Batching and Prompting.}}
LLM prompts included the task instruction, full source and target schema descriptions, target-side sample rows, and a set of up to 10-20 source queries. Prompts for query mapping were batched wherever possible to reduce API calls. 





\begin{table*}[ht]
\large
\caption{
Evaluation results on the BIRD and SPIDER benchmarks (for \textsf{Gemini-1.5-flash} and \textsf{GPT-4o-mini}). Semantic quality metrics—NL Meaningfulness and SQL–NL Alignment—are based on an LLM-as-a-judge valuation using Gemini-2.0-flash. Manual evaluation results on a smaller 144-query subset are reported in Table~\ref{tab:manual-eval}.
}
\centering
\begin{tabular}{llcccc}
\toprule
\multirow{2}{*}{\textbf{Category}} & \multirow{2}{*}{\textbf{Metric}} 
& \multicolumn{2}{c}{\textbf{BIRD Dataset}} 
& \multicolumn{2}{c}{\textbf{SPIDER Dataset}} \\
\cmidrule(lr){3-4} \cmidrule(lr){5-6}
& & \textbf{Gemini-1.5-flash} & \textbf{GPT-4o-mini} & \textbf{Gemini-1.5-flash} & \textbf{GPT-4o-mini} \\
\midrule
\multirow{3}{*}{\textit{Mapping Accuracy}} 
    & Generation Success     & 99.47\% & 100.00\% & 98.9\% & 100.00\% \\
    & Structural Alignment   & 66.59\% & 80.47\% & 82.21\% & 86.8\% \\
    & Execution Validity     & 89.43\% & 68.33\% & 93.5\% & 86.29\% \\
\midrule
\multirow{2}{*}{\textit{Semantic Quality}} 
    & NL Meaningfulness      & 95.45\% & 80.03\% & 96.41\% & 91.78\% \\
    & SQL–NL Alignment       & 82.75\% & 54.57\% & 90.02\% & 76.13\% \\
\bottomrule
\end{tabular}
\label{tab:overall-eval}
\end{table*}

\subsection{Metrics}

We report results across two dimensions: (i) mapping success and fidelity, and (ii) semantic plausibility of the generated queries.

\paragraph{\textbf{Mapping Accuracy.}}
We first measure how reliably the model can produce mapped queries under the SQL-Exchange using three metrics:

\begin{itemize}[left=1.2em, itemsep=0.3em]
  \item \textbf{Generation Success}: Percentage of queries for which the LLM generates a non-empty, parseable response instead of refusing or skipping the mapping.
  \item \textbf{Structural Alignment}: Whether the mapped query preserves the structural skeleton of the source SQL—including keywords, control blocks (e.g., \texttt{SELECT}, \texttt{JOIN}, \texttt{GROUP BY}, subqueries), and overall logic. Table and column names, constants, and optional \texttt{AS} aliasing are ignored, and basic comparison operators (e.g., \texttt{=}, \texttt{<}, \texttt{>}) are normalized.
  \item \textbf{Execution Validity}: Whether the generated SQL executes successfully on the target schema using SQLite.
\end{itemize}


\paragraph{\textbf{Semantic Quality.}}
We evaluate whether the generated queries are semantically valid via:
\begin{itemize}[left=1.2em, itemsep=0.3em]
  \item \textbf{NL Meaningfulness}: Whether the generated natural language question is clear, specific, and meaningful in the context of the target database schema.
  \item \textbf{SQL–NL Alignment}: Whether the SQL query faithfully implements the question’s intent.
\end{itemize}

\subsection{Mapping Accuracy Analysis}


Table~\ref{tab:overall-eval} summarizes aggregate results on BIRD and SPIDER, reporting mean performance for each benchmark across mapping accuracy and semantic quality. We next analyze the quality of the SQL–NL pairs produced by SQL-Exchange.

\subsubsection{\textbf{Generation Success}}

A mapping query was produced in nearly all cases, with failures---defined as explicit \textsf{null} outputs---occurring in under 1\% of instances. This behaviour varied across LLMs: \textsf{GPT-4o-mini} consistently generated a mapping for every input, whereas \textsf{Gemini-1.5-flash} occasionally returned \textsf{null} values, particularly for complex queries or when structural incompatibilities existed between the source and target schemas.


For example, one failure case involved a source query in BIRD with an invalid \textsf{HAVING} clause combining \textsf{MAX} and \textsf{MIN} without proper aggregation logic:


\begin{tcolorbox}[
  sqlbox, 
  title=Failure Case 1 (Invalid logic),
  colbacktitle=failureheader,
]

SELECT ... GROUP BY ... HAVING \textcolor{highlightred}{MAX(ibu) AND MIN(ibu)} LIMIT 2
\end{tcolorbox}


\noindent 
Here, the model correctly identified the logical flaw and refused to produce a misleading translation. 

In another case, the source query identified the employee with the most inspections in March 2016:






\begin{tcolorbox}[sqlbox, title=Failure Case 2 (Schema mismatch), colbacktitle=failureheader,]
SELECT ... FROM (

\quad SELECT employee\_id, COUNT(...) FROM ... 

\quad WHERE \textcolor{highlightred}{strftime('\%Y-\%m', ...) = '2016-03'}

\quad GROUP BY employee\_id 

\quad ORDER BY COUNT(...) DESC LIMIT 1 

) AS T2 INNER JOIN ... ON ...
\end{tcolorbox}


\noindent
This query combines temporal filtering, aggregation, ordering, and a join to retrieve personnel metadata. The model declined to generate a corresponding query for the \textsf{toxicology} schema, noting that the source logic was “difficult to directly translate to the simpler structure of the target schema.” Importantly, the \textsf{toxicology} schema contains no temporal attributes or event logs (as highlighted in red), making the transformation fundamentally infeasible. Rather than hallucinating a misleading mapping, the model conservatively returned a \textsf{null} output. This behavior reflects SQL-Exchange's robustness in recognizing when faithful transformation is not possible.

\begin{figure*}[t]
  \centering
  \includegraphics[width=.95\linewidth]{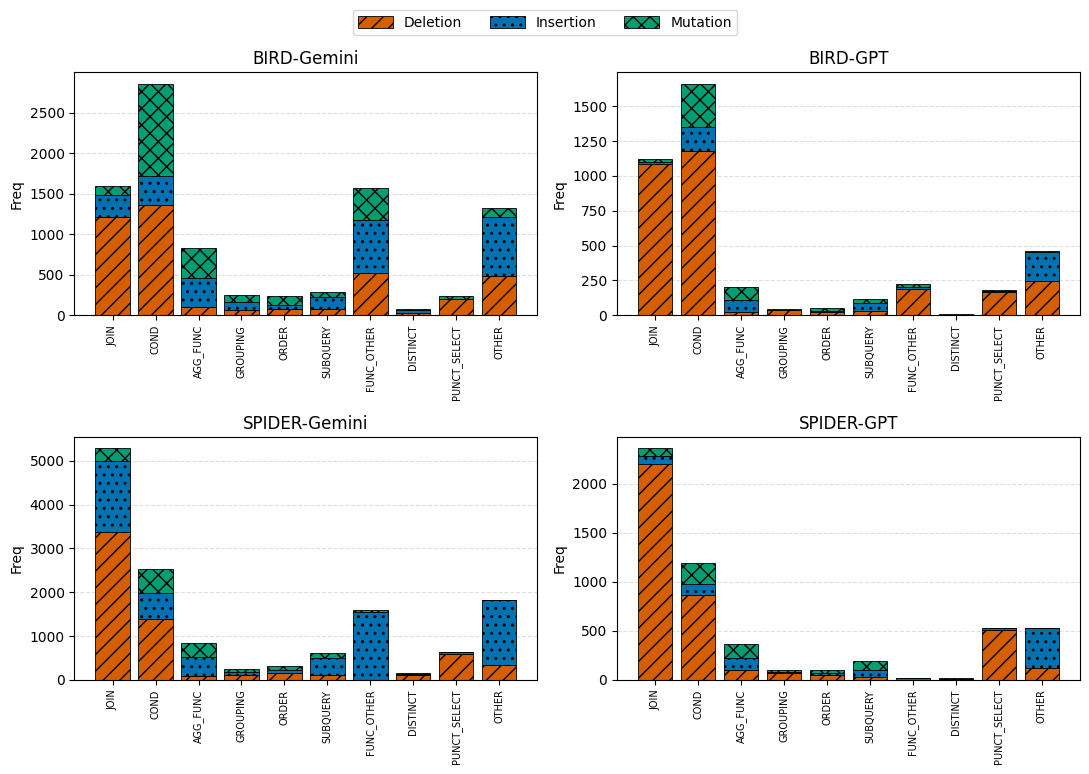}
  \caption{Frequency of structural changes, broken down to SQL constructs, 
           for the two benchmarks (\textsc{BIRD}, \textsc{SPIDER}) and two
           LLMs (\textsc{Gemini-1.5-flash}, \textsc{GPT-4o-mini}). Each bar is stacked by
           \textcolor{orange}{\bfseries deletions}, \textcolor{blue}{\bfseries insertions},
           and \textcolor{green!60!black}{\bfseries mutations}.
           Since a single query may involve multiple types of edits across different clauses, bucket totals may exceed the number of modified queries. This visualization reveals which parts of the SQL skeleton are most frequently altered during schema adaptation. Note that the charts are not on the same scale, and the dataset used for the SPIDER–Gemini setting is larger than that for SPIDER–GPT.}
      \Description{Bar charts showing the frequency of structural SQL edits made during query mapping. The edits are categorized into types such as JOIN, COND, AGG_FUNC, etc., and grouped by deletion, insertion, and mutation. Results are shown separately for BIRD and SPIDER benchmarks, and for two models: Gemini-1.5-flash and GPT-4o-mini. The visualization highlights which SQL components are most often changed during schema adaptation.}

  \label{fig:structure-diff-all}
\end{figure*}

\begin{figure*}[t]
  \centering
  \scriptsize 
  \newcommand{\examplewidth}{0.49\textwidth}

  \begin{minipage}[t]{\examplewidth}
    \begin{tcolorbox}[examplebox,
      title=\textbf{Example A: SPIDER (\texttt{architecture} $\rightarrow$ \texttt{voter\_1})},
      box align=top, valign=center]
      \begin{tcolorbox}[
        colback=orange!10, colframe=orange!60!black, boxrule=0.3pt]
        \textbf{Source NL:}
        \vspace*{0.3em}
        
        \hspace*{1.2em} How many architects haven’t built a mill before year 1850?
        \vspace*{0.2em}
        \\
        \textbf{Source SQL:}
        \vspace*{0.3em}

        \hspace*{1.2em} \sqlkw{SELECT COUNT(*) FROM} architect \\
        \hspace*{1.2em} \sqlkw{WHERE} id \sqlkw{NOT IN} \\
        \hspace*{2.5em}(\sqlkw{SELECT} architect\_id \sqlkw{FROM} mill \\
        \hspace*{2.5em}\ \sqlkw{WHERE} built\_year < 1850)

      \end{tcolorbox}

      \tcblower

      \begin{tcolorbox}[colback=teal!10, colframe=teal!60!black, boxrule=0.3pt]
        \textbf{Target NL:} 
        \vspace*{0.3em}
        
        \hspace*{1.2em} How many contestants did not receive any votes before the \hspace*{1.2em} year 2019?
        \vspace*{0.2em}
        \\
        \textbf{Target SQL:} 
        \vspace*{0.3em}
        
         \hspace*{1.2em} \sqlkw{SELECT COUNT(*) FROM} CONTESTANTS \\
         \hspace*{1.2em} \sqlkw{WHERE} contestant\_number \sqlkw{NOT IN} \\
         \hspace*{2.5em}(\sqlkw{SELECT} contestant\_number \sqlkw{FROM} VOTES \\
         \hspace*{2.5em} \sqlkw{WHERE} \uline{\sqlkw{STRFTIME}}\sqlkw{(}'\%Y', created\sqlkw{)} < '2019')

      \end{tcolorbox}
    \end{tcolorbox}
  \end{minipage}
  \hfill
  \begin{minipage}[t]{\examplewidth}
    \begin{tcolorbox}[examplebox,
      title=\textbf{Example B: BIRD (\texttt{menu} $\rightarrow$ \texttt{student\_club})},
      box align=top, valign=center, enhanced jigsaw]
      
      \begin{tcolorbox}[
        colback=orange!10, colframe=orange!60!black, boxrule=0.3pt]
        \textbf{Source NL:}
        \vspace*{0.3em}
        
        \hspace*{1.2em} For how many times had the dish ``Chicken gumbo'' \\
        \hspace*{1.2em} appeared on a menu page?
        \vspace*{0.2em}
        \\
        \textbf{Source SQL:}
        \vspace*{0.3em}
        
        \hspace*{1.2em} \sqlkw{SELECT} \sqlkw{SUM}(\sqlkw{CASE WHEN} T1.name = 'Chicken gumbo'
        \hspace*{1.2em} \sqlkw{THEN} 1 \sqlkw{ELSE} 0 \sqlkw{END}) \sqlkw{FROM} Dish AS T1 \\
        \hspace*{2.5em} \uline{\sqlkw{INNER JOIN}} MenuItem AS T2
        \uline{\sqlkw{ON}} T1.id = T2.dish\_id
      \end{tcolorbox}

      \tcblower
    
      \begin{tcolorbox}[
        colback=teal!10, colframe=teal!60!black,
         boxrule=0.3pt]
        \textbf{Target NL:} 
        \vspace*{0.3em}
        
        \hspace*{1.2em} How many times has the expense 'Water, chips, cookies' \\
        \hspace*{1.2em} been recorded? 
        \vspace*{0.2em}
        \\
        \textbf{Target SQL:} 
        \vspace*{0.3em}
        
         \hspace*{1.2em}\sqlkw{SELECT SUM} \\
         \hspace*{2.5em}\sqlkw{(CASE WHEN} T1.expense\_description = 'Water, chips, \\
          \hspace*{2.5em}\sqlkw{cookies' THEN} 1 \sqlkw{ELSE} 0 \sqlkw{END}) \\
          \hspace*{1.2em} \sqlkw{FROM} expense AS T1

      \end{tcolorbox}

    \end{tcolorbox}
  \end{minipage}

  \caption{
    Representative examples of structural edits during query mapping by SQL-Exchange.
    \textbf{(A)} Introduction of the \textsf{STRFTIME} function to enable temporal filtering on string-based timestamps;
    \textbf{(B)} Removal of a join when the target schema supports a direct column-based formulation of the counting logic.
    Examples are drawn from the SPIDER and BIRD training sets and mapped to target schemas from their respective development sets.
  }
  \Description{
    Two side-by-side examples illustrating structural edits made by SQL-Exchange during schema mapping.
    Example A adapts a query from SPIDER's architecture schema to voter\_1 by replacing a numeric year filter on a date column with a STRFTIME-based filter on a timestamp field.
    Example B adapts a BIRD query by removing a join and instead using a direct column reference in the target schema to perform conditional counting.
    Each example includes the source natural language question and SQL query, followed by the mapped target question and SQL.
  }
  \label{fig:structure-examples}
\end{figure*}

\subsubsection{\textbf{Structural Alignment}} 
In 66\% to 87\% of cases, the mapped queries accurately preserved the structural skeleton of the source SQL queries while adapting to the target schema. Perfect structural preservation is not always feasible due to differences between source and target schemas, which may require modifying joins,  groupings, or aggregation functions to ensure semantic alignment.

To better understand how LLMs adapt query structures during mapping, we categorize token-level edits into ten semantic buckets based on SQL keywords: \textsf{JOIN} (join clauses), \textsf{COND} (filters and logical conditions), \textsf{DISTINCT} (keyword), \textsf{FUNC\_OTHER} (scalar, formatting, and null-handling functions), \textsf{AGG\_FUNC} (aggregate functions), \textsf{GROUPING} (grouping and filtering), \textsf{ORDER} (ordering and row limits), \textsf{SUBQUERY} (nested queries and set operations), \textsf{PUNCT\_SELECT} (commas in \textsf{SELECT} clause indicating changes in the number of output columns), and \textsf{OTHER} (residual edits). Each change is labeled as a deletion, insertion, or mutation, and may fall into multiple categories. For instance, editing a fragment like \textsf{JOIN ... WHERE ... IN (SELECT ...)} may trigger \textsf{JOIN}, \textsf{COND}, and \textsf{SUBQUERY} simultaneously.

Figure~\ref{fig:structure-diff-all} presents the distribution of these edits for Gemini-1.5-flash and GPT-4o-mini on BIRD and SPIDER. Deletions dominate, followed by insertions and fewer mutations. This trend suggests that when LLMs modify query structure, they often simplify queries by removing joins, filters, or modifiers that do not directly align with the target schema.

We observe that the majority of structural edits occur in the \textsf{JOIN} and \textsf{COND} buckets. These include deletions of unnecessary joins or filters and mutations that generalize conditions (e.g., transforming `=` to `LIKE`). While the model frequently simplifies joins, it also introduces conditions or filters to adapt queries to the target schemas. Gemini tends to perform more insertions in \textsf{FUNC\_OTHER}, \textsf{AGG\_FUNC}, and \textsf{SUBQUERY}, such as adding \textsf{STRFTIME()}, \textsf{LENGTH()}, or \textsf{COUNT()} to improve compatibility with target schema semantics. Similar patterns are observed in GPT-4o-mini mappings, though Gemini tends to perform more structural substitutions involving functions and aggregations. 

SPIDER queries exhibit a stronger tendency toward deletion, particularly in \textsf{JOIN} and \textsf{COND} categories. This is likely due to the higher degree of normalization and the presence of auxiliary tables in SPIDER schemas. These characteristics can make certain joins redundant or overly complex in the mapped queries.

In terms of overall change patterns, deletions account for 55.2\% of structural edits, followed by insertions (32.2\%) and mutations (12.6\%). These statistics reinforce the observation that LLMs simplify query structure when faced with schema mismatches, even though the overall skeleton is often preserved.


To complement the aggregate statistics, Figure~\ref{fig:structure-examples} presents representative examples of structural edits made during query mapping. These illustrate SQL-Exchange’s ability to adapt queries to diverse schemas: (A) introducing temporal functions to bridge column-type mismatches, such as adapting a year comparison (\texttt{built\_year < 1850}) to a timestamp field using \texttt{strftime('\%Y', created) < '2019'}; and (B) removing joins when the target schema provides a direct column that captures the same semantic condition, allowing the query to be expressed more simply. These edits reflect the model’s schema-sensitive and structure-aware behavior.





\subsubsection{\textbf{Execution Validity}}
As shown in Table~\ref{tab:overall-eval}, execution success rates for mapped queries range from 68.3\% to 93.5\%, across models and datasets, with \textsf{Gemini-1.5-flash} performing best. This indicates that, in most cases, the mapped queries are both syntactically correct and semantically well-aligned with the target schema. 


To better understand the failure cases, we categorized all SQL error messages into six classes. Most errors are \textbf{column reference errors} (90.5–93.4\%), typically involving incorrect table–column associations, nonexistent columns, or missing quotes for identifiers containing spaces. A smaller portion consists of \textbf{syntax errors} (1.9–6.3\%)—such as missing commas, stray parentheses, or unescaped identifiers—which appear more frequently in \textsf{GPT-4o-mini} outputs. Other errors include \textbf{table reference mistakes} (0.3–3.2\%) from incorrect or hallucinated table names; \textbf{ambiguous column names} (0.7–2.5\%), which occur when join-related columns lack disambiguation; and \textbf{misuses of aggregate functions} (0.2–2.5\%), such as invalid or nested \texttt{SUM}, \texttt{MAX}, or \texttt{COUNT}. The remaining \textbf{rare errors} ($\leq$1.2\%) involve malformed set operators, incomplete \texttt{SELECT} clauses, improperly nested subqueries, or timeouts caused by complex joins or missing indexes. This breakdown shows that even structurally correct queries can fail due to subtle schema mismatches, suggesting that future work may benefit from schema-level validation or improved quoting mechanisms.

\subsubsection{\textbf{Structure vs. Utility}}


While structural alignment is a key goal of our method, Table~\ref{tab:overall-eval} shows that preserving structure alone does not guarantee better performance. GPT-4o-mini exhibits higher structural alignment than Gemini-1.5-flash on both benchmarks, yet performs significantly worse in execution validity and semantic alignment, especially on BIRD. This suggests that strict structural preservation can hurt semantic consistency or produce non-executable SQL. In contrast, Gemini-1.5-flash makes small structural adjustments when needed, producing more executable queries that better align with the natural language intent. These findings indicate that effective cross-schema mapping requires balancing structural fidelity with semantic adaptability.

\noindent\subsubsection{\textbf{Open-Weight Model Evaluation.}}
While Table~\ref{tab:overall-eval} reports full-benchmark results only for \textsf{Gemini-1.5-flash} and \textsf{GPT-4o-mini}, we also conducted a brief open-weight model evaluation using \textsf{Llama-3.3-70B}. 
Since BIRD is substantially more complex than SPIDER, this evaluation was limited to the same BIRD subset used for the Gemini and GPT models. On this subset, Llama achieved competitive performance across all metrics (98.51\% generation success, 85.23\% structural alignment, 70.93\% execution validity, 85.55\% NL meaningfulness, and 63.54\% SQL--NL alignment). These results demonstrate that SQL-Exchange is effective under modern open-weight models as well.


\subsection{Semantic Quality of Generated Queries}

To evaluate the semantic quality of the mapped queries, we assess whether the generated natural language questions are meaningful within the context of the target schema, and whether the corresponding SQL queries are logically consistent with the intent of the generated question. This evaluation is performed both manually and through an LLM-based review.

\subsubsection{\textbf{Manual Evaluation}}


We conducted a manual evaluation on a curated subset of 144 SQL–NL pairs that passed execution testing. To ensure diverse coverage, we selected three target databases that exhibited low execution success rates or weak structural alignment in our earlier analyses: \textsf{california\_schools} and \textsf{european\_football\_2} (from BIRD), and \textsf{poker\_player} (from SPIDER). For each, we randomly sampled eight source schemas and selected three structurally varied queries per pair—categorized as simple ($\leq$5 elements), moderate (6–9), or complex ($\geq$ 10), based on the number of tables and columns. This produced 72 mappings, evaluated under two LLMs (\textsf{GPT-4o-mini} and \textsf{Gemini-1.5-flash}), yielding 144 total examples.


Each of the three authors independently judged whether (1) the NL question was meaningful for the target schema and (2) the SQL query matched that question’s intent. Only cases with agreement from at least two annotators were included—resulting in 138 judgments for NL quality and 134 for SQL–NL alignment.



As shown in Table~\ref{tab:manual-eval}, across all settings, the vast majority of generated questions (100\% for Gemini-1.5-flash and 88-91\% for GPT-4o-mini) were judged meaningful. SQL–NL alignment was high for \textsf{Gemini-1.5-flash} (91–96\%) and moderate for \textsf{GPT-4o-mini} (56–57\%), highlighting a notable gap in fidelity across models.

\begin{table}[t]
\caption{
Manual evaluation of 144 mapped queries, assessing NL meaningfulness and SQL-NL alignment. Annotator agreement was reached at 138 queries for NL meaningfulness and 134 for SQL–NL alignment. Trends align with LLM-based results in Table~\ref{tab:overall-eval}.
}
\large
\centering
\begin{tabular}{lcc}
\toprule
\textbf{Benchmark-Model} & \textbf{Meaningful NL} & \textbf{Aligned SQL} \\
\midrule
BIRD–Gemini   & 100.0\% & 91.1\% \\
BIRD–GPT   & 88.6\%  & 57.1\% \\
SPIDER–Gemini & 100.0\% & 95.8\% \\
SPIDER–GPT & 91.3\%  & 56.5\% \\
\bottomrule
\end{tabular}

\label{tab:manual-eval}
\end{table}

\subsubsection{\textbf{Validating LLM-as-a-Judge Against Human Annotations}} 

To assess the potential of LLMs as scalable evaluators, we used \textsf{Gemini-2.0-flash} to score the same 144 query pairs evaluated manually. As with the manual evaluation, we report results only for the subset of queries where at least two out of three annotators agreed: 138 for NL meaningfulness and 134 for SQL–NL alignment. \textsf{Gemini-2.0-flash} achieved \textbf{94.9\%} agreement on question meaningfulness and \textbf{85.07\%} on SQL–NL alignment, closely matching human annotations. Most disagreements were either borderline “maybe” cases, due to subjective interpretation. In the full disagreements (e.g., one labeled as “yes” and the other as “no”), the LLM’s explanation was often more consistent and grounded than the human rationale, suggesting that LLM-based evaluation can serve as a reliable alternative to manual review. 

\subsubsection{\textbf{Scaling Up with LLM-as-a-Judge}}

After \textsf{Gemini-2.0-flash} was validated against human judgment, we used it to evaluate the full set of mapped queries. As shown in Table~\ref{tab:overall-eval}, most generated NL questions were judged meaningful (95–96\% for \textsf{Gemini-1.5-flash} and 80–91\% for \textsf{GPT-4o-mini}). SQL–NL alignment was similarly strong for \textsf{Gemini-1.5-flash} (82–90\%), and more variable for \textsf{GPT-4o-mini} (54–76\%), reflecting a consistent performance gap across models and datasets.
Importantly, this evaluation was conducted on the complete set of generated queries, regardless of whether they passed structural alignment or execution validity checks. This allows us to assess the semantic plausibility even in structurally or syntactically imperfect outputs, which may explain the slightly lower scores compared to the manually curated subset.

\subsection{Ablation: SQL-Exchange vs. Zero-Shot}

To assess the contribution of each component in our methodology, we compare \textsc{SQL-Exchange} with a zero-shot prompting baseline. Both approaches use the same LLMs and schema information, but the baseline omits key elements of our framework: structured Chain-of-Though, structural query templates, demonstration examples, and target schema sample rows. 
This comparison isolates the impact of structural abstraction and schema grounding on mapping quality.


We evaluate both approaches using the core metrics introduced in Table~\ref{tab:overall-eval}, and additionally introduce a new metric, \textbf{Result Yield}, which measures whether the generated query returns non-empty results on the target database. This metric complements execution validity by assessing whether the query is not only syntactically correct but also semantically grounded in the target schema's data.

\begin{figure}[t]
  \centering
    \includegraphics[width=\linewidth]{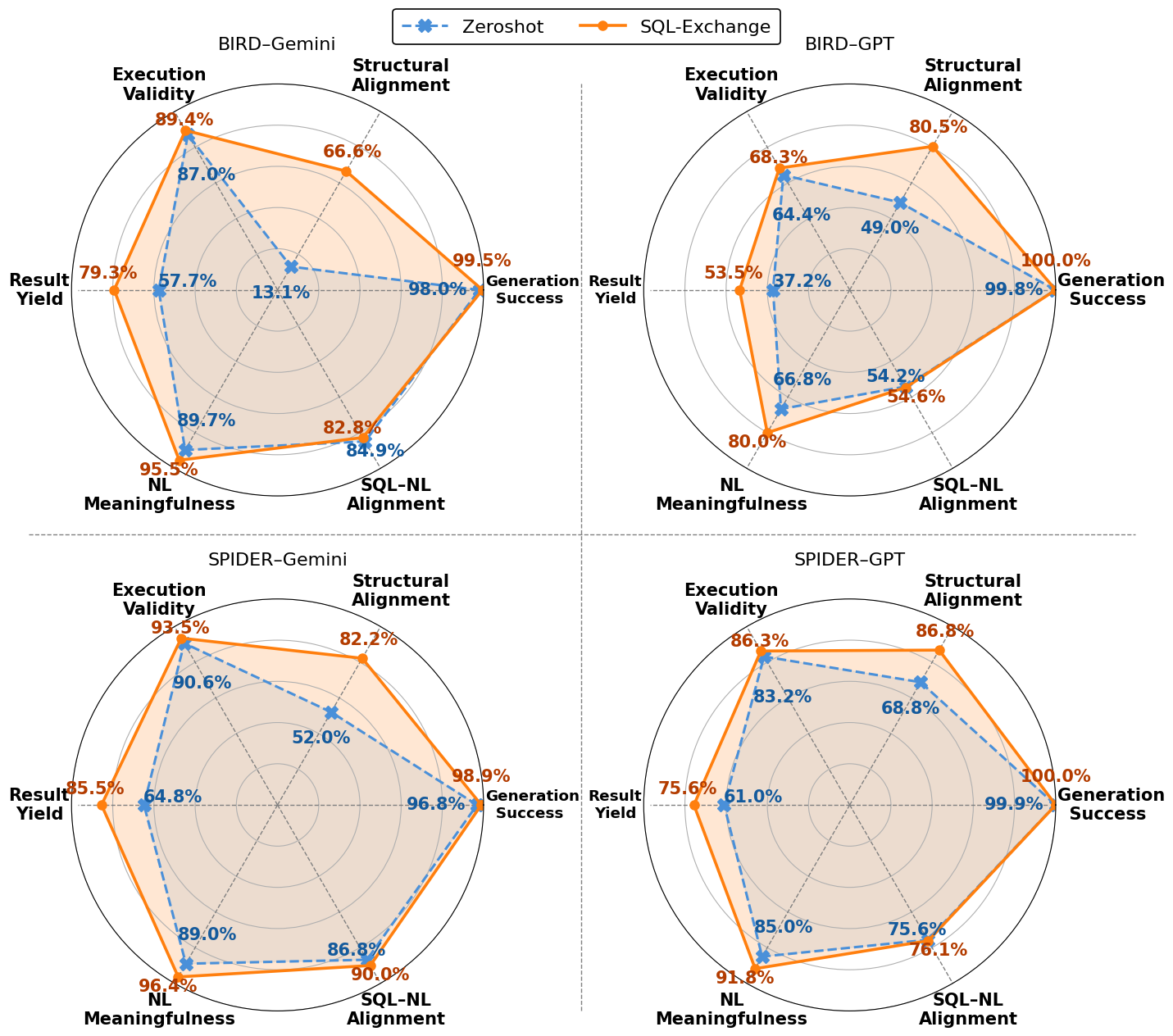}
  \caption{Comparison of SQL-Exchange and Zero-Shot prompting across semantic and structural metrics.}
  \Description{
    Four radar plots comparing SQL-Exchange and zero-shot prompting across six metrics:
    structural alignment, SQL–NL alignment, NL meaningfulness, execution validity, result yield, and generation success.
    Each subplot corresponds to a different setting: BIRD–Gemini, BIRD–GPT, SPIDER–Gemini, and SPIDER–GPT.
    In all cases, the orange region (SQL-Exchange) encloses or exceeds the blue region (zero-shot), showing consistent gains across metrics.
  }
  \label{fig:radar-bird}
\end{figure}


Figure~\ref{fig:radar-bird} shows four radar plots comparing SQL-Exchange and zero-shot prompting across six evaluation metrics, one plot for each model–dataset setting. SQL-Exchange demonstrates consistent gains in both structural and semantic quality: structural alignment improves by +18\% to +53.53\%, result yield increases by +14.7\% to +21.7\%, and meaningful questions rates improve by +5.3\% to +13.3\%. SQL-Exchange also consistently outperforms zero-shot on both \textit{Generation Success} and \textit{Execution Validity} across all settings, underscoring the effectiveness of SQL-Exchange’s abstraction-driven templates and schema-grounded constant substitution mechanisms.

Performance on \textit{NL–SQL alignment} is more mixed. Zero-shot has a slight advantage in the BIRD-Gemini setting, but SQL-Exchange performs better in the remaining three. Even when alignment scores appear similar, zero-shot queries often simplify the structure by omitting essential components such as \textsf{JOIN} clauses or conditional logic. Although the resulting queries may remain executable or logically aligned in intent, they diverge from the original compositional semantics. In contrast, SQL-Exchange preserves structural scaffolding far more faithfully.

We further analyze SQL-Exchange's behavior relative to the zero-shot baseline along three dimensions: (1) structural simplification through deletions, (2) preservation of relational joins, and (3) avoidance of undesirable reuse from the source schema.


\paragraph{\textbf{Structural simplification via deletion.}}
 On average, zero-shot prompting exhibited significantly more deletions than the mappings produced by SQL-Exchange---ranging from approximately 1.5× to 4.5× depending on the model and benchmark. The most frequently deleted components were JOIN clauses and conditional filters, which are essential for preserving multi-table logic and row-level constraints. This indicates a strong simplification trend, where zero-shot prompting omits critical structural elements rather than faithfully adapting them.


\begin{figure}[t]
  \centering
  \includegraphics[width=0.95\linewidth]{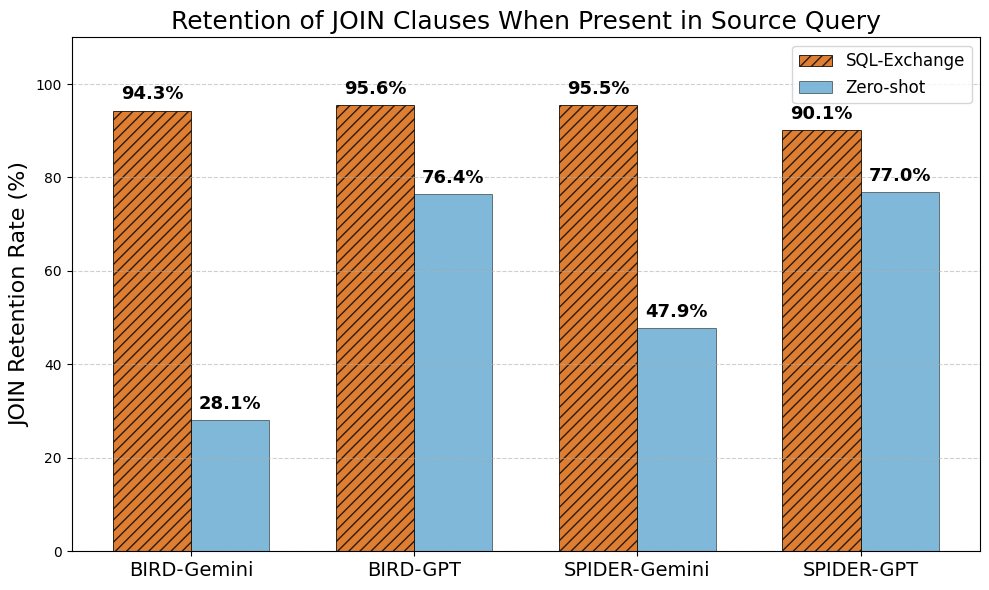}
  \caption{Join retention rates: the proportion of mapped queries that preserve a \textsf{JOIN} present in the source. SQL-Exchange retains joins consistently ($>$90\%), while zero-shot often drops them, especially with Gemini.}
  
  \Description{
    Bar chart comparing JOIN clause retention rates between SQL-Exchange and zero-shot prompting for four settings: 
    BIRD–Gemini, BIRD–GPT, SPIDER–Gemini, and SPIDER–GPT. 
    SQL-Exchange shows consistently high retention rates (above 90\%) across all settings, 
    while zero-shot prompting performs worse, especially in BIRD–Gemini (28.1\%) and SPIDER–Gemini (47.9\%). 
    Bars are color-coded with orange for SQL-Exchange and blue for zero-shot, and numeric retention rates are displayed above each bar.
  }
  \label{fig:join-retention}
\end{figure}

\begin{figure*}[t]
  \centering
  \includegraphics[width=\linewidth]{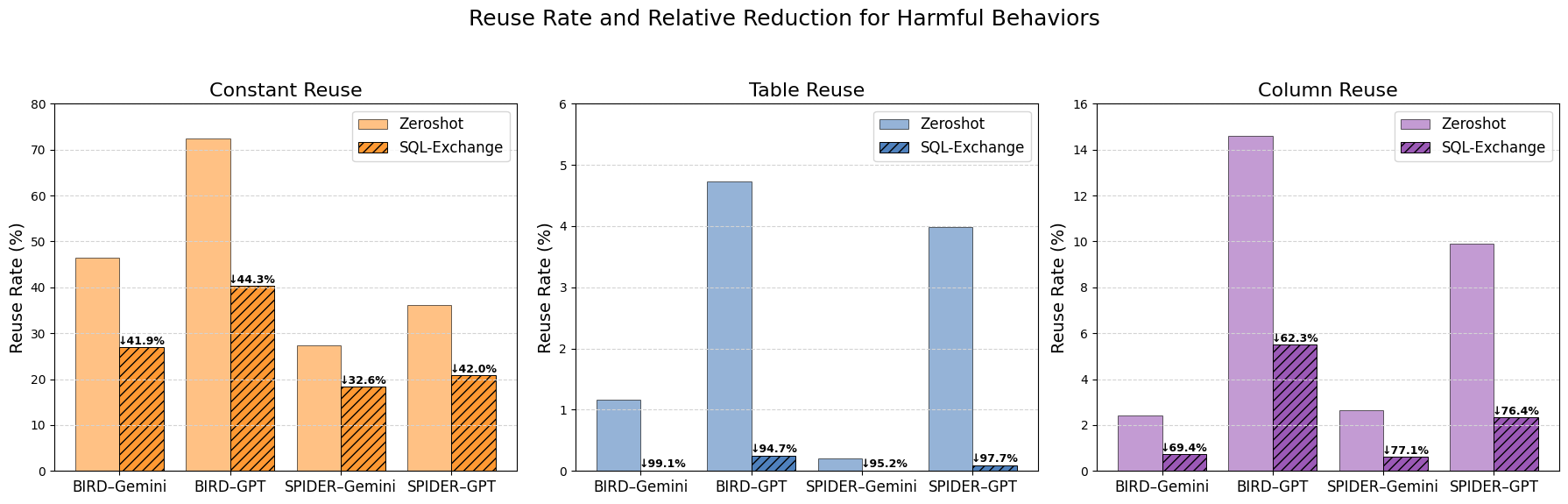}
  \caption{Reduction in schema-specific reuse across settings. 
  We compare reuse rates for constants, tables, and columns between zero-shot prompting and SQL-Exchange. Each subplot shows actual reuse percentages across four schema settings, with SQL-Exchange in solid bars and zero-shot in transparent bars. Above each SQL-Exchange bar, we annotate the \textbf{relative reduction} (\%) compared to the zero-shot baseline. SQL-Exchange dramatically lowers the rate of constant, table, and column reuse.}
  \Description{
    A three-panel bar chart comparing reuse rates for constant reuse, table reuse, and column reuse across four schema settings: BIRD-Gemini, BIRD-GPT, SPIDER-Gemini, and SPIDER-GPT.
    Each panel shows two sets of bars: transparent bars for zero-shot prompting and solid bars for SQL-Exchange.
    The constant reuse panel shows that SQL-Exchange reduces reuse from 46\% to 27\% in BIRD-Gemini and from 72\% to 40\% in BIRD-GPT.
    The table reuse panel shows drastic reductions, such as from 4.8\% to 0.3\% in BIRD-GPT.
    The column reuse panel shows drops from 15\% to 5.6\% in BIRD-GPT and from 10\% to 2.4\% in SPIDER-GPT.
    Each bar includes a downward arrow and a percent decrease to indicate relative reduction.
  }
  \label{fig:harmful-behaviors}
\end{figure*}

\paragraph{\textbf{Preservation of relational joins.}}
 As shown in Figure~\ref{fig:join-retention}, zero-shot prompting often fails to retain \textsf{JOIN} clauses in the mapped output when the source query originally included one, resulting in structurally oversimplified outputs. On BIRD-Gemini and SPIDER-Gemini, fewer than 30\% and 50\% of such queries, respectively, contain any join clause after mapping. In contrast, SQL-Exchange retains joins in over 90\% of cases across all settings, suggesting that schema abstraction in SQL-Exchange enables more faithful multi-table reasoning when adapting query structure across domains.

\paragraph{\textbf{Avoidance of undesirable schema reuse.}}
Figure~\ref{fig:harmful-behaviors} shows that SQL-Exchange substantially reduces the reuse of constants, tables, and columns from the source schema—an issue that plagues zero-shot prompting. Table reuse decreases by over 90\% relative to the zero-shot baseline in all settings, falling below 0.5\%. Column reuse drops by 62–77\% relative to zero-shot, demonstrating consistent suppression of schema-copying artifacts. This confirms that SQL-Exchange's schema-aware design not only preserves structural logic more reliably, but also minimizes unintended copying of irrelevant source-specific elements.

\section{Downstream Task Performance}
\label{sec:text-to-sql}

Beyond structural and semantic correctness, we also assess whether the mapped queries are useful as in-context examples or fine-tuning data in downstream text-to-SQL tasks. The key question is whether these schema-aligned queries improve SQL generation accuracy in one-shot/few-shot prompting or when fine-tuning local models.

\subsection{Experimental Setup}

We evaluate three instruction-tuned open-source models: \textsf{Llama-3.2-3B-Instruct}, \textsf{Qwen2.5-Coder-3B-Instruct}, and \textsf{Qwen2.5-Coder-7B-Instruct}~\citep{hui2024qwen25}. In addition, for supervised fine-tuning experiments, we use \textsf{Mistral-7B-Instruct-v0.3} and \textsf{CodeLlama-7b-Instruct}. Throughout the paper, these models are referred to as \texttt{LLaMA-3B}, \texttt{Qwen-3B}, \texttt{Qwen-7B}, \texttt{Mistral-7B}, and \texttt{CodeLlama-7B}, respectively. We also include results from the proprietary models \textsf{GPT-4o-mini} and \textsf{GPT-4o}.





\paragraph{\textbf{Inference.}}  
Open-source models were run locally with Hugging Face Transformers on a single NVIDIA A100 (40GB) using \textsf{float16}. Inference used greedy decoding (\textsf{temperature = 0.0}, \textsf{top\_p = 1.0}) and instruction-style formatting.  
Fine-tuning was performed on Google Colab with an 80GB A100.
\textsf{GPT} models were accessed via the OpenAI API with the same greedy decoding setup.


\paragraph{\textbf{In-Context Example Selection.}} 

We evaluated multiple strategies for selecting in-context examples, including 
 embedding-based retrieval, masked-question selection,  DAIL~\citep{gao2024text}, and SQL-Encoder\footnote{\url{https://huggingface.co/MrezaPRZ/sql-encoder}}. SQL-Encoder is a 1.3B-parameter model trained to estimate structural similarity between SQL queries using only their natural language questions~\citep{pourreza2024sql}. For all selection methods, candidates were retrieved under two standard settings: (i) from source-side queries (unmapped) and (ii) from mapped outputs produced by SQL-Exchange. To obtain an approximate upper bound on achievable performance, we additionally used SQL-Encoder to retrieve examples from the target development sets as well. Retrieval was performed independently for each model–dataset setting and filtered by semantic and execution validity where applicable.



\subsection{Metrics}

We measure downstream performance using \textbf{execution accuracy}, the standard text-to-SQL metric (percentage of generated queries matching the gold execution result). To analyze quality control during example selection, we also report \textbf{filtering ablations}, removing semantic and execution filters to isolate their contribution. Semantic filtering removes NL–SQL mismatches, while execution filtering excludes non-executable queries.

\begin{table*}[t]
\begin{threeparttable}
\caption{Execution accuracy (EX) on BIRD and SPIDER development sets across LLMs and example-selection methods. Results are reported as BIRD / SPIDER. 
SQL-Exchange examples consistently improve over zero-shot and unmapped baselines, and SQL-Encoder with oracle retrieval provides an approximate upper bound. 
}
\label{tab:text_to_sql_combined}
\centering
\small
\setlength{\tabcolsep}{6.5pt}
\begin{tabular}{llccccc}
\toprule
\multirow{2}{*}{\textbf{Selection Method}} & \multirow{2}{*}{\textbf{Prompting Strategy}} &
\textbf{LLaMA-3B} &
\textbf{Qwen-3B} &
\textbf{Qwen-7B} &
\textbf{GPT-4o-mini} &
\textbf{GPT-4o} \\
& &
\textbf{BIRD/SPIDER} &
\textbf{BIRD/SPIDER} &
\textbf{BIRD/SPIDER} &
\textbf{BIRD/SPIDER} &
\textbf{BIRD/SPIDER} \\
\midrule

\textit{No In-Context Examples} 
  & Zero-shot
  & 26.53 / 43.33
  & 35.98 / 53.48
  & 49.80 / 71.08
  & 50.46 / 66.73
  & 56.39 / 69.34 \\

\midrule

\multirow{2}{*}{\textit{Embedding-based}} 
  & Unmapped (1-shot)
  & 22.88 / 40.52
  & 32.66 / 52.80
  & 47.26 / 62.57
  & 50.39 / 67.21
  & 58.67 / 70.70 \\
  & SQL-Exchange (1-shot)
  & 32.40 / 52.61
  & 39.37 / 60.15
  & 50.52 / 67.22
  & 52.48 / 72.15
  & 58.21 / 73.98 \\

\midrule
\multirow{2}{*}{\textit{Masked Question}} 
  & Unmapped (1-shot)
  & 23.60 / 42.94
  & 33.57 / 53.38
  & 46.94 / 64.41
  & 49.54 / 69.92
  & 58.60 / 71.66 \\
  & SQL-Exchange (1-shot)
  & 31.68 / 52.13
  & 39.70 / 59.28
  & 50.39 / 66.34
  & 52.35 / 69.83
  & 59.58 / 73.98 \\

\midrule
\multirow{3}{*}{\textit{DAIL~\citep{gao2024text}}}

  & Unmapped$^{\dagger}$ (1-shot)
  & 23.92 / 42.75
  & 34.68 / 56.58
  & 48.63 / 65.57
  & 50.33 / 69.63
  & 59.19 / 71.76 \\
  & SQL-Exchange$^{\dagger}$ (1-shot)
  & 31.42 / 50.97
  & 38.40 / 59.19
  & 49.35 / 65.96
  & 51.24 / 67.50
  & 58.02 / 71.86 \\
  & Unmapped$^{\Diamond}$ (1-shot)
  & 24.05 / 46.23
  & 35.53 / 58.90
  & 47.00 / 65.76
  & 51.04 / 72.73
  & 60.69 / 75.43 \\
  & SQL-Exchange$^{\Diamond}$ (1-shot)
  & 33.64 / 53.97
  & 42.11 / 63.83
  & 49.48 / 69.34
  & 52.28 / 73.79
  & 59.91 / 77.66 \\

\midrule
\multirow{6}{*}{\textit{SQL-Encoder}} 

  & Unmapped (1-shot)
  & 25.03 / 43.42
  & 31.94 / 60.93
  & 46.94 / 70.12
  & 48.37 / 68.86
  & 58.74 / 70.60 \\
  & SQL-Exchange (1-shot)
  & 35.33 / 57.64
  & 41.46 / 70.70
  & 51.89 / 73.60
  & 51.50 / 69.73
  & 58.67 / 68.28 \\
  & Unmapped (3-shot)
  & 20.73 / 37.62
  & 34.29 / 53.97
  & 46.87 / 71.66
  & 49.15 / 69.44
  & 60.50 / 72.63 \\
  & SQL-Exchange (3-shot)
  & 32.66 / 56.67
  & 43.09 / 68.67
  & 50.85 / 73.89
  & 53.19 / 73.40
  & 60.17 / 78.05 \\
  & Oracle (1-shot)
  & 39.05 / 76.02
  & 48.89 / 84.82
  & 58.21 / 90.14
  & 57.11 / 84.04
  & 63.10 / 80.56 \\
  & Oracle (3-shot)
  & 34.94 / 74.66
  & 49.15 / 87.52
  & 59.65 / 90.04
  & 58.60 / 86.17
  & 65.78 / 86.26 \\

\bottomrule

\end{tabular}

\begin{tablenotes}[flushleft]
\footnotesize
\item[$\dagger$] DAIL performed using \textsf{GPT-4o-mini} for query similarity selection.
\item[$\Diamond$] DAIL performed using the gold SQL query as an oracle for query similarity selection.
\end{tablenotes}

\end{threeparttable}

\end{table*}


\subsection{Prompting Strategies and Results}


We evaluate a broad set of prompting strategies, grouped by how in-context examples are selected: \textbf{Zero-shot}, which generates SQL without examples; \textbf{Embedding-based (1-shot)}, which retrieves the closest example using Sentence Transformers\footnote{We use the \texttt{all-mpnet-base-v2} model for embedding-based retrieval.}; and \textbf{Masked Question (1-shot)}, which retrieves examples using a masked-question representation of the NL input. 
We also include \textbf{DAIL (1-shot)}~\citep{gao2024text}, which selects examples via Query Similarity Selection; to do this, we use either a preliminary model (\textsf{GPT-4o-mini}, $^{\dagger}$) or the gold SQL as an oracle ($^{\Diamond}$) to identify the best example from each candidate pool. 
Finally, we use \textbf{SQL-Encoder (1-shot / 3-shot)}, which retrieves structurally similar examples using NL-based structural similarity, and we apply it in both 1-shot and 3-shot settings as well as in an oracle mode using target dev-set pairs.

Table~\ref{tab:text_to_sql_combined} reports execution accuracy on BIRD and SPIDER across all models and selection methods. As seen throughout the table, unmapped examples frequently underperform even zero-shot, especially for smaller models, whereas SQL-Exchange consistently provides stronger demonstrations that better transfer across schemas.

Across nearly all retrieval methods and models, SQL-Exchange substantially outperforms Unmapped examples. On SPIDER, gains frequently exceeding +5 EX and reaching as high as +19.05 EX for LLaMA-3B under SQL-Encoder (3-shot). On BIRD, improvements similarly reach up to +11.93 EX.
For GPT-4o, improvements are smaller in a few settings, not because SQL-Exchange is ineffective, but because demonstrations were generated using substantially weaker models (Gemini-1.5-flash and GPT-4o-mini). We expect larger gains when SQL-Exchange examples are produced using a model comparable in strength to GPT-4o. Given that most SQL-Exchange vs. Unmapped differences exceed the natural variance of SPIDER ($\approx$1–1.5 EX) and BIRD ($\approx$1.5–2 EX), paired significance tests confirm that the majority of SQL-Exchange improvements are statistically significant (typically $p < 0.05$, often $p < 0.01$), with only very small deltas (<2 EX) falling in the nonsignificant range.

In the SQL-Encoder setting, both 1-shot and 3-shot methods generally outperform their unmapped counterparts when using SQL-Exchange.
Although we apply 3-shot prompting only within SQL-Encoder, results show that adding more schema-aligned examples helps larger models (e.g., +5.4\% for GPT-4o on SPIDER), while smaller models sometimes see mild declines due to added noise.


Overall, SQL-Exchange provides effective and robust in-context examples across model scales, retrieval methods, and prompting strategies, without requiring task-specific fine-tuning or large synthetic corpora.




\paragraph{\textbf{Filtering Ablations}} Table~\ref{tab:text_to_sql_ablation_combined} examines the effect of removing execution or semantic filtering when selecting mapped examples. 
We restrict our ablation study to the SQL-Encoder setting, which supports both 1-shot and 3-shot prompting and offers the most complete retrieval pipeline, making it a representative case for isolating filter effects.
Extending ablations to all methods and models would require substantially more LLM calls, which is impractical in time and cost. For the same reason, \textsf{GPT-4o} is not included.


Across models, removing semantic filtering causes the largest accuracy drops—especially for smaller LMs (e.g., –1.4\% on BIRD for \textsf{LLaMA-3B}, –1.36\% on SPIDER for \textsf{Qwen-3B}). This shows that keeping the NL question aligned with the mapped SQL is crucial. Execution filtering has a smaller impact and can even help slightly when removed for stronger models (e.g., +0.5\% for \textsf{GPT-4o-mini}), indicating that larger LMs can tolerate minor SQL imperfections. Smaller models, however, are more sensitive to both types of errors, underscoring the value of filtering for lower-capacity settings.



\begin{table}[t]
\caption{Ablation study showing the effect of removing execution or semantic filtering on accuracy for both BIRD and SPIDER development sets in one-shot and few-shot settings. Semantic filtering contributes most to downstream performance across both prompting strategies.}
\centering
\begin{tabular}{llcc}
\toprule
\textbf{Model} & \textbf{Configuration} & \textbf{BIRD} & \textbf{SPIDER} \\
\midrule
\multirow{6}{*}{\textsf{LLaMA-3B}} 
  & SQL-Exchange (1-shot)              & 35.33\% & 57.64\% \\
  & \quad w/o execution filter         & 34.03\% & 57.83\% \\
  & \quad w/o semantic filter          & 33.96\% & 55.61\% \\
  & SQL-Exchange (3-shot)           & 32.66\%  & 56.67\% \\
  & \quad w/o execution filter         & 32.20\%  & 56.87\% \\
  & \quad w/o semantic filter          & 32.33 \%  & 53.87\% \\
\midrule
\multirow{6}{*}{\textsf{Qwen-3B}} 
  & SQL-Exchange (1-shot)              & 41.46\% & 70.7\% \\
  & \quad w/o execution filter         & 40.48\% & 70.6\% \\
  & \quad w/o semantic filter          & 40.74\% & 69.34\% \\
  & SQL-Exchange (3-shot)           & 43.09\%  & 68.67\% \\
  & \quad w/o execution filter         & 41.26\%  & 69.25\% \\
  & \quad w/o semantic filter          & 42.05\%  & 67.12\% \\
\midrule
\multirow{6}{*}{\textsf{Qwen-7B}} 
  & SQL-Exchange (1-shot)              & 51.89\% & 73.6\% \\
  & \quad w/o execution filter         & 51.17\% & 73.4\% \\
  & \quad w/o semantic filter          & 50.78\% & 72.24\% \\
  & SQL-Exchange (3-shot)           & 50.85\%  & 73.89\% \\
  & \quad w/o execution filter         & 50.39\%  & 73.6\% \\
  & \quad w/o semantic filter          & 50.39\%  & 73.69\% \\
\midrule
\multirow{6}{*}{\textsf{GPT-4o-mini}} 
  & SQL-Exchange (1-shot)              & 51.5\%  & 69.73\% \\
  & \quad w/o execution filter         & 51.56\% & 70.21\% \\
  & \quad w/o semantic filter          & 52.74\% & 70.21\% \\
  & SQL-Exchange (3-shot)           & 53.19\%  & 73.4\% \\
  & \quad w/o execution filter         & 53.39\%  & 74.27\% \\
  & \quad w/o semantic filter          & 52.93\%  & 74.47\% \\
\bottomrule
\end{tabular}

\label{tab:text_to_sql_ablation_combined}
\end{table}

\paragraph{\textbf{Understanding the Oracle–Mapped Gap}} While our method consistently outperforms zero-shot and unmapped one-shot baselines, it still falls short of Oracle performance. This gap arises because Oracle examples, drawn from the development set, often closely mirrors the target query logic—sharing joins, filters, ordering, and aggregation. In contrast, SQL-Exchange mappings are constrained by the availability and structure of source queries, which may lack strong logical or schema-level alignment.

Mapped queries sometimes retain redundant joins (e.g., auxiliary tables) or introduce irrelevant filters, misleading the LLM. Oracle examples also employ more precise constructs (e.g., \textsf{IS NOT NULL}, \textsf{DISTINCT}) and exact aggregation logic that better match the NL question. Lastly, Oracle queries often reflect more natural phrasing and keyword overlap with the test NL question, further helping the model follow the correct reasoning path. 


\subsection{Finetuning}


To further assess the utility of SQL-Exchange, we fine-tune downstream models on two of the most challenging BIRD development databases—\textit{california\_schools} and \textit{toxicology}—where baseline performance is lowest. Following the setup of~\citep{pourreza2024dts}, we fine-tune \textsf{Mistral-7B} on two datasets: (1) our 1.3K sampled source-database queries from the BIRD training set, and (2) 1.3K SQL-Exchange–generated mappings created by transferring those queries into the target schema. On \textit{california\_schools}, zero-shot \textsf{Mistral-7B} achieves \textbf{6.74\%} execution accuracy; fine-tuning on source-only data reduces accuracy to \textbf{5.62\%}, whereas fine-tuning on SQL-Exchange–generated data boosts it to \textbf{14.61\%}. 
A similar trend appears on \textit{toxicology}, with accuracy improving from \textbf{8.28\%} (zero-shot) to \textbf{10.34\%} (source-only) and \textbf{22.07\%} with SQL-Exchange–generated data.

Similar behavior is observed with \textsf{CodeLlama-7B}. On \textit{california\_schools}, accuracy moves from \textbf{3.37\%} zero-shot to \textbf{1.12\%} with source-only fine-tuning, but rises to \textbf{7.87\%} when fine-tuned on SQL-Exchange mappings. On \textit{toxicology}, performance increases from \textbf{8.97\%} to \textbf{15.17\%} with source-only fine-tuning and to \textbf{20.00\%} with SQL-Exchange data. These results demonstrate that SQL-Exchange provides substantially more effective supervision on difficult target schemas than source-only data.

\section{Conclusion}

We introduced a novel schema-aware method for mapping SQL queries across database schemas using LLMs, a problem not directly addressed in prior work. Our approach preserves the structural skeleton of the source query while adapting schema-specific elements and constants through one-shot prompting, template-based abstraction, and sample-guided substitution. Our evaluations on BIRD and SPIDER show that most of the mapped queries are structurally aligned with the source, executable on the target, and semantically consistent with the generated natural language. Furthermore, when used as in-context examples in downstream text-to-SQL tasks, these mappings improve execution accuracy over zero-shot and unmapped baselines, and fine-tuning on SQL-Exchange–generated mappings provides substantial additional gains on challenging databases compared to fine-tuning on unmapped queries or relying on zero-shot prompting.


\begin{acks}
This research was supported by the Natural Sciences and Engineering Research Council of Canada (NSERC). Computational resources were provided by the Digital Research Alliance of Canada.
\end{acks}


\bibliographystyle{ACM-Reference-Format}
\bibliography{myrefs}



\end{document}